\documentclass[9pt, sigconf, authorversion, nonacm]{acmart}

\setcopyright{none}
\usepackage{subcaption}
\usepackage{algpseudocode}
\usepackage{algorithm}

\algnewcommand\algorithmicforeach{\textbf{for each}}
\algdef{S}[FOR]{ForEach}[1]{\algorithmicforeach\ #1\ \algorithmicdo}

\newcommand{\name}{R-HLS}

\begin{document}

\title{\name{}: An IR for Dynamic High-Level Synthesis and \\ Memory Disambiguation based on Regions and State Edges}

\author{David Metz}
\email{David.C.Metz@NTNU.no}
\orcid{0000-0001-7103-7968}
\affiliation{%
  \institution{Norwegian University of Science and Technology (NTNU)}
  \streetaddress{Department of Computer Science}
%  \city{Trondheim}
  \country{}
%  \postcode{7491}
}

\author{Nico Reissmann}
\email{Nico.Reissmann@gmail.com}
\orcid{0000-0002-4096-1821}
\affiliation{%
  \institution{Independent Researcher}
  \country{}
}

\author{Magnus Sj\"alander}
\email{Magnus.Sjalander@NTNU.no}
\orcid{0000-0003-4232-6976}
\affiliation{%
  \institution{Norwegian University of Science and Technology (NTNU)}
  \streetaddress{Department of Computer Science}
%  \city{Trondheim}
  \country{}
%  \postcode{7491}
}


\begin{abstract}

Dynamically scheduled hardware enables high-level synthesis~(HLS) for applications with irregular control flow and latencies, which perform poorly with conventional statically scheduled approaches.
Since dynamically scheduled hardware is inherently data flow based, it is beneficial to have an intermediate representation~(IR) that captures the global data flow to enable easier transformations.
State-of-the-art dynamic HLS utilize control flow based IRs, which model data flow only at the basic block level, requiring the rediscovery of inter-block parallelism.
The Regionalized Value State Dependence Graph (RVSDG) is an IR that models (1) control flow as part of the global data flow utilizing regions and (2) memory dependencies using state edges.

We propose \name{}, a new RVSDG dialect targeted for dynamic high-level synthesis.
\name{} explicitly models control flow decisions, routing, and memory, which are only abstractly represented in the RVSDG.
% Contributions
Expressing the control flow as part of the data flow reduces the need for complex optimizations to extract performance and enables easy conversion to parallel circuits.
Furthermore, we present a distributed memory disambiguation optimization that leverages memory state edges to decouple address generation from data accesses, resulting in resource efficient out-of-program-order execution of memory operations.
%The state edges enable fine-grained out-of-order execution of memory accesses without external load-store queues.
% Results

Our results show that \name{} effectively exposes parallelism, resulting in fewer executed cycles and a 10\% speedup on average, compared to the state-of-the-art in dynamic HLS with optimized memory disambiguation.
These results are achieved with a significant reduction in resource utilization, such as a 79\% reduction in lookup-tables and 22\% reduction in flip-flops, on average.

% Conclusion

%    Dynamic scheduling enables efficient high-level synthesis (HLS) 

%    We present an intermediate representation (IR) that is ``ideally'' suited for high-level synthesis (HLS) that is based on regions with dataflow, state edges, and local memory.

 %   We show how this IR easily maps onto dynamically scheduled circuits (also called elastic circuits) and how the IR enables the creation of scalable dynamic memory disambiguation for out-of-order memory accesses.
\end{abstract}

%\begin{CCSXML}
%<ccs2012>
%  <concept_id>00000000.0000000.0000000</concept_id>
%  <concept_desc>Do Not Use This Code, Generate the Correct Terms for Your Paper</concept_desc>
%  <concept_significance>500</concept_significance>
%  <concept_id>00000000.00000000.00000000</concept_id>
%  <concept_desc>Do Not Use This Code, Generate the Correct Terms for Your Paper</concept_desc>
%  <concept_significance>300</concept_significance>
%  <concept_id>00000000.00000000.00000000</concept_id>
%  <concept_desc>Do Not Use This Code, Generate the Correct Terms for Your Paper</concept_desc>
%  <concept_significance>100</concept_significance>
%  <concept_id>00000000.00000000.00000000</concept_id>
%  <concept_desc>Do Not Use This Code, Generate the Correct Terms for Your Paper</concept_desc>
%  <concept_significance>100</concept_significance>
%</ccs2012>
%\end{CCSXML}

%\ccsdesc[500]{Do Not Use This Code~Generate the Correct Terms for Your Paper}
%\ccsdesc[300]{Do Not Use This Code~Generate the Correct Terms for Your Paper}
%\ccsdesc{Do Not Use This Code~Generate the Correct Terms for Your Paper}
%\ccsdesc[100]{Do Not Use This Code~Generate the Correct Terms for Your Paper}

%\keywords{HLS, RVSDG}

\maketitle
\thispagestyle{plain}
\pagestyle{plain}

\section{Introduction}

Dynamically scheduled hardware enables the schedule of operations to adapt to dynamic events, such as data dependent control flow and memory access latencies.
This is similar to how an out-of-order (OoO) scheduled core adapts to dynamic events in contrast to an in-order (InO) scheduled core where the compiler relying on static information dictates the schedule.
The recent interest in dynamic high-level synthesis (HLS)~\cite{dynamically-scheduled-hls:FPGA2018, dynamatic:FPGA2020, fluid:ASYNC2021, fast-tokens:FPL2022, inter-block-scheduling:FPL2022, parallel-control-flow:TRETS2023, straight-to-the-queue:FPGA2023} 
%--- which transforms application code to dynamically scheduled hardware --- 
is due to the generated dynamically scheduled hardware outperforming conventional static HLS, which struggles with efficiently transforming irregular application code.

Dynamically scheduled hardware consists of components that communicate through a handshaking protocol.
When all incoming signals are marked active, the component consumes the inputs and produces one or more outputs, which in turn signals dependent components that a new input is available.
% Might want to say something about rate limitation, i.e., the new signaling of the new value is blocked until the previous output value has been consumed by the dependent components.
The goal of a dynamic HLS tool is to extract as much parallelism as possible by enabling as many components as possible to execute at the same time.
The available parallelism in an application is fundamentally limited by its data and memory dependencies.
Data dependencies are the values that flow directly between components, while memory dependencies are memory operations that (might)\footnote{The HLS tool has to rely on static information to perform alias analysis, making it impossible to determine if two memory operations alias or not in general.} access the same memory location.

All HLS tools must adhere to these dependencies to ensure correct functional behavior.
%
%State-of-the-art dynamic HLS tools~\cite{dynamatic:FPGA2020, dynamically-scheduled-hls:FPGA2018, fluid:ASYNC2021, parallel-control-flow:TRETS2023} are based on basic blocks, where the data flow is separated between intra (within the) block and inter (between) block data flow, which complicates data and memory dependence analyses, as well as transformations.
%\\ Alternatively: \\
State-of-the-art dynamic HLS tools~\cite{dynamically-scheduled-hls:FPGA2018, dynamatic:FPGA2020, fluid:ASYNC2021, fast-tokens:FPL2022, inter-block-scheduling:FPL2022, parallel-control-flow:TRETS2023, straight-to-the-queue:FPGA2023} are based on a control data flow graph (CDFG), where dependencies within a basic block (\textit{intra block}) are modeled as a data flow graph while the basic blocks are connected in a control flow graph that represents \textit{inter block} dependencies.
The separation between intra and inter block dependencies complicates data and memory dependence analyses, as well as optimizations.
Since the task of the HLS tool is to identify all data and memory dependencies, it should be beneficial to have an intermediate representation~(IR) that explicitly encodes all data and memory dependencies as a global data flow graph.

The Regionalized Value State Dependence Graph (RVSDG)~\cite{rvsdg:TECS2020} is a data-flow based IR that models control flow as regions and memory dependencies as state edges.
RVSDG has been shown to effectively encode all data and memory dependencies of complete applications in a single graph on which optimizations are performed~\cite{rvsdg:TECS2020}.
RVSDG's global data flow and state edges consistently encode both intra and inter region dependencies, but the current RVSDG implementation is targeted for conventional compilation to machine instructions intended for execution on a general purpose processor.
The current RVSDG~\cite{rvsdg:TECS2020} represents many details only at an abstract level, e.g., memory operations are modeled but not the memory on which they operate, details which are required for performing HLS.

In this paper, we present \name{} an RVSDG dialect for dynamic HLS.
\name{} maintains the core properties of RVSDG while being more expressive to capture details that are fundamental for HLS.
Our main contributions are:
\begin{itemize}
    \item We demonstrate the suitability of RVSDG for HLS and create a new RVSDG dialect called \name{} that enables the generation of efficient dynamically scheduled hardware (\autoref{sec:rhls-dialect}).
%    \item We extend RVSDG with support for explicitly representing memory. 
    \item We exploit memory state edges to create distributed and efficient dynamic memory disambiguation (\autoref{sec:memory-disambiguation}).
    \item We show that \name{}'s global data flow representation effectively exposes parallelism, resulting in circuits with consistently low cycle counts (\autoref{sec:results}) without the need for complex optimizations.
%    \item We show that \name{}'s global data flow representation exposes more parallelism of the application, eliminating or significantly simplifying transformations needed by CDFG based representations to generate parallel hardware.
\end{itemize}

Our results, from a set of irregular applications intended for dynamic HLS, show that \name{} exposes more parallelism, resulting in consistently fewer executed cycles and, on average, a 10\% reduction in execution time, compared to the state-of-the-art in dynamic HLS with optimized memory disambiguation called ``Straight to the Queue'' by A. Elakhras et al.~\cite{straight-to-the-queue:FPGA2023}.
The results also show that \name{} in combination with our proposed memory disambiguation is resource efficient, resulting in a 79\% reduction in look-up tables and 22\% reduction in flip-flops.

\section{Background}
\label{sec:background}

In this section, we give a short introduction to dynamically scheduled hardware and the Regionalized Value State Dependence Graph (RVSDG) intermediate representation.

\subsection{Dynamically Scheduled Hardware}
\label{sec:dynamic-hardware}

Conventional high-level synthesis commonly uses a finite state machine to schedule operations~\cite{introHLS:IEEEDT2009, legup:FPGA2011}.
This works well when all latencies are statically known and the various paths in a circuit are of similar length, or the achievable performance is determined by a single path, e.g., when the optimal start of the next iteration of a loop can be statically predetermined.
The resulting hardware implementations are efficient, as a limited amount of logic is required to implement the control flow.
However, when latencies are unknown at compile time, which is the case for irregular code that has variable length memory accesses or data dependent control flow, static HLS must conservatively assume the worst case latency, often resulting in poor performance.
Statically scheduled hardware behaves analogous to VLIW (very long instruction word) processors, where each instruction determines what operations to be executed in parallel, as defined by the compiler.
If a single operation within a VLIW instruction experiences a longer latency than assumed by the compiler, then the whole instruction must stall, blocking all forward progress.

Dynamically scheduled hardware (also called elastic circuits~\cite{elastic-architectures:DAC2006, elastic-systems:MEMOCODE2010, fast-tokens:FPL2022}) is inspired by asynchronous systems and out-of-order execution, where an operation is performed as soon as its inputs are available.
This is achieved through a handshake protocol that is used to communicate data values between two components.
Each data signal is augmented with two single-bit handshake control signals, one that signals to the dependent operation that a new data value is available and one (going the opposite direction) that signals that the dependent operation has consumed the data value.
A commonly used terminology is that of tokens, where a component generates a token that is passed with the data and that a dependent component can consume.

\looseness -1
We describe the most common components that normally are implemented to support dynamically scheduled hardware.
The handshake protocol only supports point-to-point communication between two distinct components.
Fan-out, i.e., multiple operations being dependent on a single produced value, is supported by a dedicated \textit{fork} component, which takes one input and replicates it a predefined number of times as individual outputs.
Each output can then be connected to one of the dependent components.
Control flow is implemented as dedicated components, such as \textit{branch} and \textit{merge}.
The branch takes a single input and propagates an input token to one of its outputs as specified by a condition.
Branch is used to implement if, switch, and loop statements.
Merge is the opposite of branch.
It has many inputs, and a token on an input is directly propagated to its single output.
Merge is used for converging control flow, similar to phi-nodes in static single assignment form.
Data values are routed using the \textit{select} (also called \textit{multiplexer}) component, which takes two or more inputs and based on a select signal routes the input to the output, while discarding the incoming tokens on the non-selected inputs.
\textit{Buffer} and \textit{FIFO} components are used for inserting D~flip-flops and FIFO queues, which can hold token(s).
FIFOs enable, e.g., short producer paths (paths with few/short-latency operations) to run ahead of longer consumer paths (paths with many/long-latency operations).
The shorter path writes into the FIFO, enabling it to continue its execution instead of waiting for the longer path to consume the token.
This decouples the timing of the two paths and enables them to execute out of order.

The purpose of dynamic HLS is to perform efficient translation of irregular applications into dynamically scheduled hardware, enabling HLS for a class of applications that previously have resulted in poor performance.

\subsection{Regionalized Value State Dependence Graph}
\label{sec:rvsdg}

The intermediate representation (IR) in a compiler is the core data structure used for representing source code, and for analyzing and transforming the code.
The expressiveness of the IR and the match between the IR representation and the final target dictates how easy it is to perform analyses and transformations.

The Regionalized Value State Dependence Graph (RVSDG)~\cite{rvsdg:TACO2015, rvsdg:TECS2020} is a data-flow centric IR that encodes data dependencies as edges between nodes, which represents operations.
Control flow is implicitly represented by so-called structural nodes, which contain regions.
Structural nodes consist of:
\textit{Omega} nodes, which have a single region that represents the whole translation unit;
\textit{Lambda} nodes, which have a single region that represents the body of a function;
\textit{Theta} nodes, which  have a single region that represents the body of a do-while loop; and
\textit{Gamma} nodes, which have multiple regions, where each region represents one branch of an if-then-else or switch statement.

A structural node maps its inputs to arguments at the top of the region and the region results to the structural node's outputs.
The arguments drive the input values of the data flow contained in the region, and the outcome of the data flow graph are connected to the region results.
% Theta
For the theta node (loops), the inputs are mapped to the arguments at the initiation of the loop, for each new loop iteration the results gets mapped back to the arguments (implicit back edges), and upon exiting the loop,  the results of the region get mapped to the outputs.
% Gamma
For the gamma node (if-then-else and switch statements), the inputs get mapped to the argument in the region that gets selected (the active region).
The results of the active region are mapped to the outputs of the node.

The RVSDG ensures correct ordering of operations with side effects by using I/O and memory state edges.
I/O edges are used to ensure the order of operations with externally visible side effects, while memory state edges are used to ensure ordering of memory operations accessing the same memory locations.
All memory operations are initially sequentialized by a single memory state edge that goes through each and every memory operation.
As memory operations are determined to be independent, they can be separated by introducing new memory state edges.
The use of state edges combined with regions containing data-flow graphs enables RVSDG to effectively represent all data and memory dependencies in one single graph.

The representation of both data and memory dependencies in a single global data-flow graph presents a unique opportunity for analyzing and transforming source code to dynamically scheduled hardware, as shown in the following sections.

\section{The \name{} Dialect}
\label{sec:rhls-dialect}

\looseness -1
RVSDG's data-flow centric representation, which captures data and memory dependencies in a single graph, makes it a strong contender for the use as an intermediate representation (IR) for high-level synthesis (HLS).
However, the current implementation of RVSDG~\cite{rvsdg:TECS2020, jlm:GITHUB2024} is targeted for conventional compilers that produce machine instructions to be executed on a general purpose processor.
Machine instructions are intended for a specific architecture (e.g., RISC-V, Arm, x86), which constrains the design space and enables an abstract representation.
For example, there is no need to model routing of data values, as this is automatically performed by reading and writing architectural registers, and there is no notion of memory ports and memory arbitration\footnote{VLIW is one exception where the number of memory ports are hard-coded into the instruction format.}.
We call the current RVSDG implementation the \textit{R-LLVM} dialect as the starting point and end result are LLVM IR~\cite{llvm:CGO2004}, and most RVSDG operations closely correspond to those found in the LLVM IR.
The abstract representation of the R-LLVM dialect makes it unsuitable for performing analyses and transformations required for performing HLS.

We present \name{}, an RVSDG dialect designed for HLS.
\name{} maintains all the core properties of RVSDG --- i.e., data dependencies are represented as a data flow graph, memory dependencies as state edges, and control flow as regions --- while being more expressive to capture details that are fundamental for HLS.

\subsection{The \name{} Design}
\label{sec:rhls-design}

The handshake protocol is implicitly modeled as part of the existing data-flow edges.
The individual edges are instantiated as a handshake bundle, consisting of ready, valid, and data signals, when \name{} is lowered to FIRRTL~\cite{firrtl:ICCAD2017}.
In R-LLVM, state edges represent the order in which machine instructions are to be generated, and have no physical manifestation.
This is in stark contrast to \name{}, where a state edge represents the control signals of the handshake protocol and physical transfer of tokens (without data).

While many simple operations, such as logical and arithmetic operations, are unchanged from the R-LLVM dialect, \name{} introduces several new types of operations.

The standard set of operations for dynamically scheduled hardware enables explicit representation of control flow, such as branch (BRANCH), (non-)discarding multiplexers (NDMUX and DMUX), and a simple buffer (BUF) (see \autoref{sec:gamma-lowering} for examples of their use).
We introduce a novel loop operation (HLS-LOOP), which replaces the R-LLVM theta node and exposes the required routing and control logic for generating hardware.
The new HLS-LOOP together with two new buffers (PRED-BUF and LOOP-BUF) are used to lower R-LLVM theta nodes to \name{} (\autoref{sec:theta-lowering}).

Memory operations (loads and stores) in \name{} have dedicated inputs and outputs for communicating with memory, and memory ports are explicitly modeled.
New specialized operations are added for handling memory requests (MEM-REQ) and responses from memory (MEM-RESP) (\autoref{sec:memory-ports}).

New address queue (ADDR-Q) and state gate (SG) operations are introduced, which enable dynamic memory disambiguation.
\name{}'s encoding of memory dependencies with state edges enables efficient implementations of dynamic and distributed memory disambiguation (\autoref{sec:memory-disambiguation}).
Dynamic memory disambiguation is a core functionality for enabling memory operations to execute out-of-program-order and is a crucial functionality for efficient dynamic HLS.

%\hms{
%Nodes mentioned in the following sections that should be introduced here: \\
%Branch (BRANCH) \\
%Discarding multiplexer (DMUX) \\
%Non-discarding multiplexer (NDMUX) \\
%HLS-LOOP \\
%Predicate buffer (PRED-BUF) \\
%Loop constant buffer (LOOP-BUF) \\
%Memory request (MEM\_REQ) \\
%Memory response (MEM\_RESP) \\
%State gate (SG) \\
%Address queeu (ADDR-Q) \\
%}

%DMUX — The discarding multiplexer discards a token from each of the inputs that are not selected by the predicate.

%HLS-LOOP — Are similar to R-LLVM nodes, but are more flexible and more explicit regarding routing of variables.
%The functionality of the HLS-LOOP and the loop specific operations PRED-BUF and LOOP-BUF are described in \autoref{sec:theta-lowering}.

%MEM\_REQ and MEM\_RESP is used for accessing memory and is described in \autoref{sec:memory-ports}.

%ADDR-Q and SQ is used for memory disambiguation, which is described in \autoref{sec:memory-disambiguation}.

%\hms{Nodes used (we probably want to mention the most important additions): \\
%* NDMUX \\
%* DMUX \\
%* Branch \\
%* Fork \\
%* CFork \\
%* SinkOp \\
%* PredicateBuffer \\
%* LoopConstantBuffer \\
%* Buffer \\
%* PassthroughBuffer \\
%* LoopNode \\
%* StateGate \\
%* AddrQueue \\
%* MemReq/MemRes \\
%* Load/Store \\
%}
%single node source - no ambiguity of origin
%ready valid semantics and propagation?

\subsection{Lowering from R-LLVM to \name{}}
\label{sec:rllvm-to-rhls}

This section describes how the R-LLVM dialect can be lowered into the proposed \name{} dialect.

\subsubsection{Lowering Gamma}
\label{sec:gamma-lowering}

Gamma nodes are lowered in two ways, depending on the data-flow graph in their subregions.
In both conversion methods, the contents of the subregions are flattened into the region containing the gamma.

\begin{figure}[t]
    \centering
    \begin{subfigure}[b]{0.2\textwidth}
        \centering
        \includegraphics[width=\textwidth]{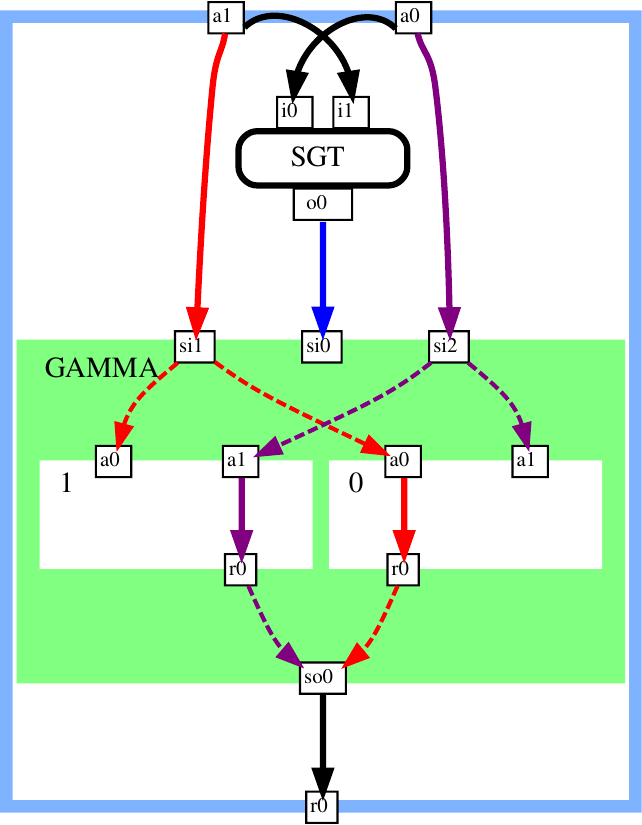}
        \caption{Gamma in R-LLVM}
        \label{fig:llvm-gamma-without-mem-state}
    \end{subfigure}
    \hfill
    \begin{subfigure}[b]{0.2\textwidth}
        \centering
        \includegraphics[width=0.5\textwidth]{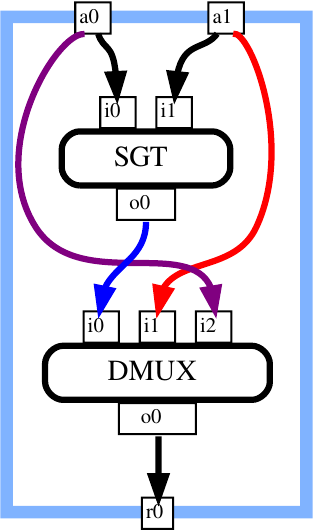}
        \caption{Converted to \name{}}
        \label{fig:rhls-gamma-without-mem-state}
    \end{subfigure}
    \vspace{-1em}
    \caption{Lowering gamma node without state edges or theta}
    \label{fig:gamma-without-mem-state}
    \vspace{-1em}
\end{figure}

If no state edges pass through the gamma and the subregions do not contain any theta nodes, then each input is directly connected to its associated argument in each subregion.
An example of this for a function that returns the bigger of its two inputs is shown in \autoref{fig:llvm-gamma-without-mem-state}.
The associated results from each region are connected via a discarding multiplexer (DMUX), which routes the result from the selected subregion to the output.
The intuition behind this lowering is that, since there are no side effecting operations (no state edges) and no complicated control flow (no loops), it is possible to execute all regions in parallel and simply select the correct result(s) by routing them to the output(s).

\begin{figure}[t]
    \centering
    \begin{subfigure}[b]{0.24\textwidth}
        \centering
        \includegraphics[width=\textwidth]{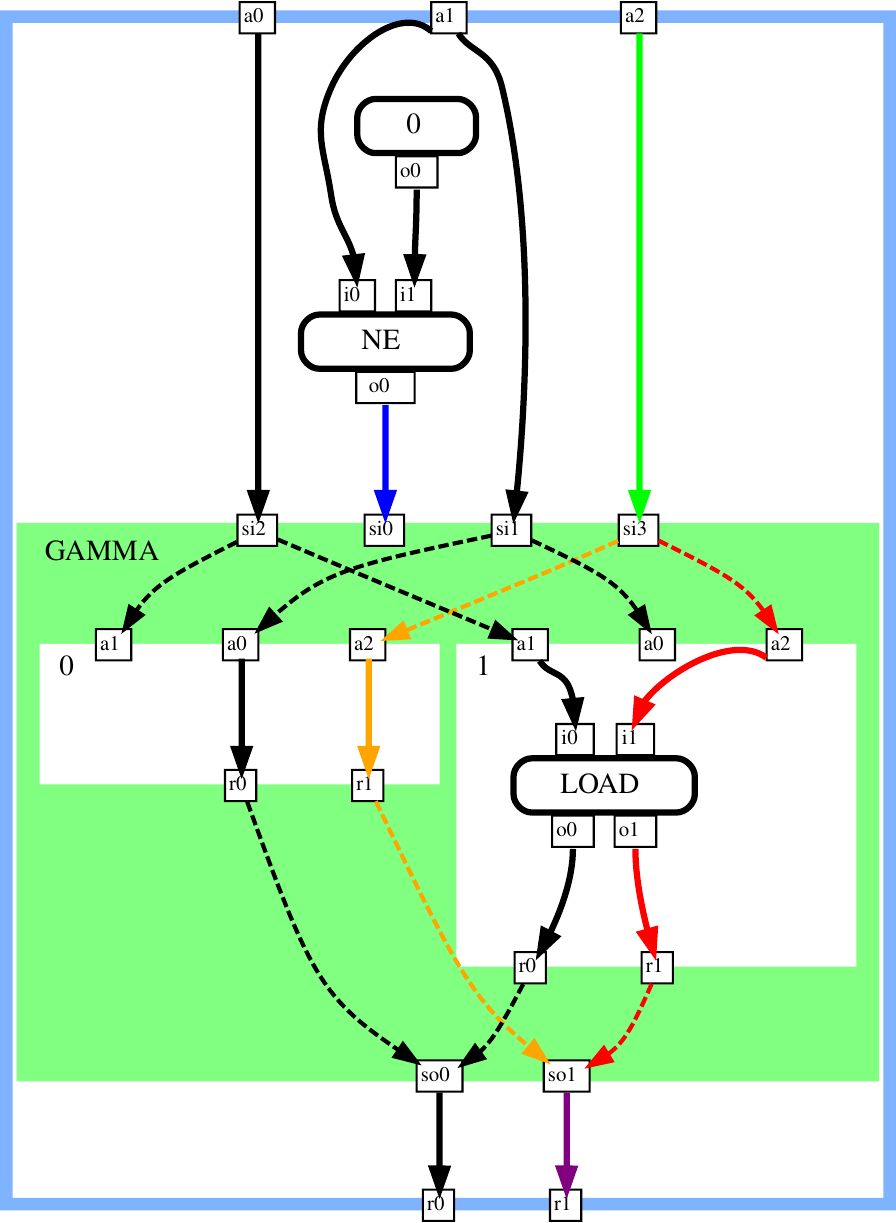}
        \caption{Gamma in R-LLVM}
        \label{fig:llvm-gamma-with-mem-state}
    \end{subfigure}
    \hfill
    \begin{subfigure}[b]{0.22\textwidth}
        \centering
        \includegraphics[width=\textwidth]{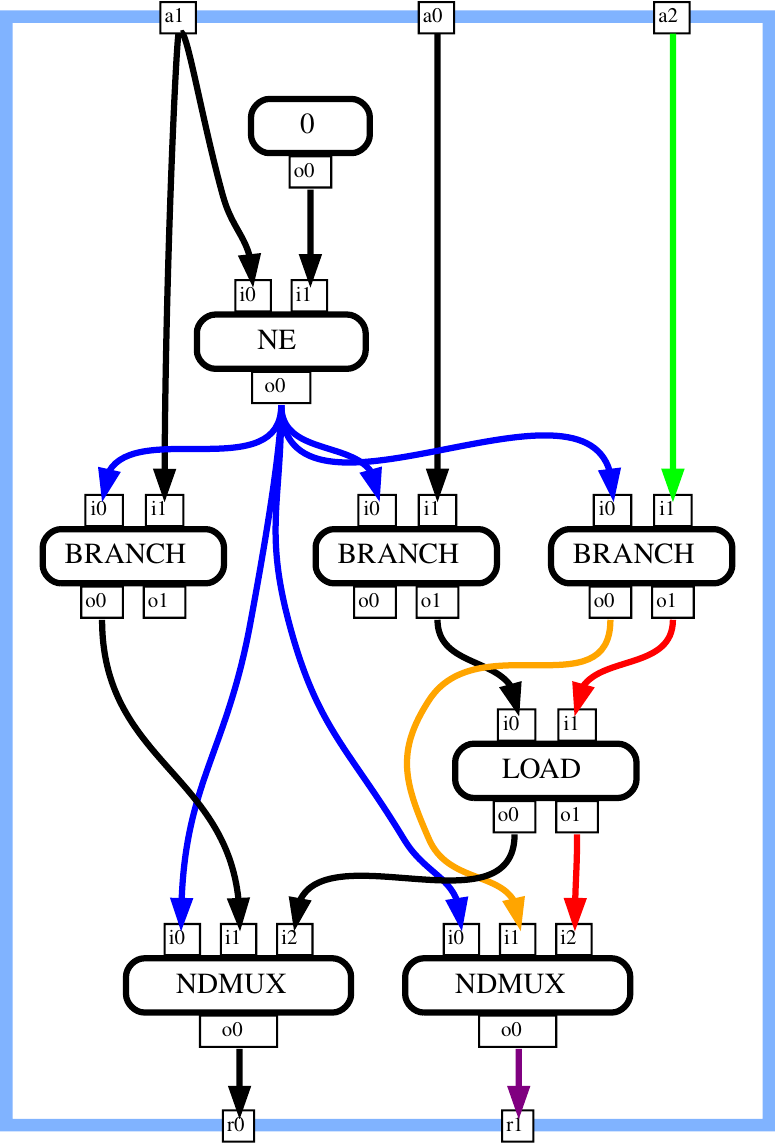}
        \caption{Lowered to \name{}}
        \label{fig:rhls-gamma-with-mem-state}
    \end{subfigure}
    \vspace{-1em}
    \caption{Lowering of gamma node with state edges}
    \label{fig:gamma-with-mem-state}
    \vspace{-1em}
\end{figure}

If a state edge passes through the gamma or a subregion contains a theta node, see \autoref{fig:llvm-gamma-with-mem-state}, then a branch operation is placed between each input of the gamma and the associated argument in each subregion.
The associated results from each region are connected via a non-discarding multiplexer (NDMUX) to the output.
The branch ensures that only the operations in the selected region are performed, and since only one region generates results, there is no need to discard the results from the other regions.

In either case, the branches and multiplexers are controlled by the predicate signal (blue) of the gamma node, which is determined by the signed greater than (SGT) and not equal (NE) node in the examples shown in \autoref{fig:gamma-without-mem-state} and \ref{fig:gamma-with-mem-state}, respectively.

\subsubsection{Lowering Theta}
\label{sec:theta-lowering}

\begin{figure}[t]
    \centering
    \begin{subfigure}[b]{0.15\textwidth}
        \centering
        \includegraphics[width=\textwidth]{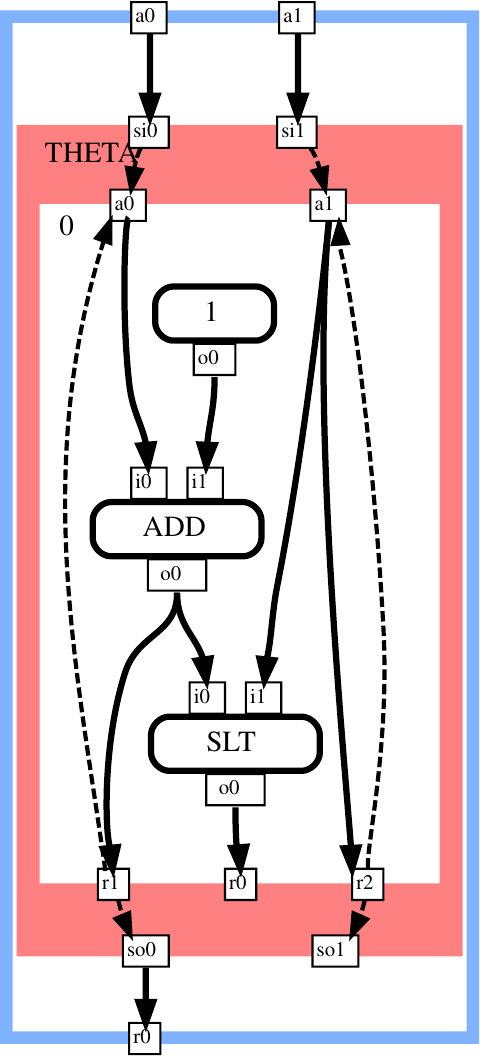}
        \caption{Theta in R-LLVM}
        \label{fig:llvm-theta}
    \end{subfigure}
    \hfill
    \begin{subfigure}[b]{0.25\textwidth}
        \centering
        \includegraphics[width=\textwidth]{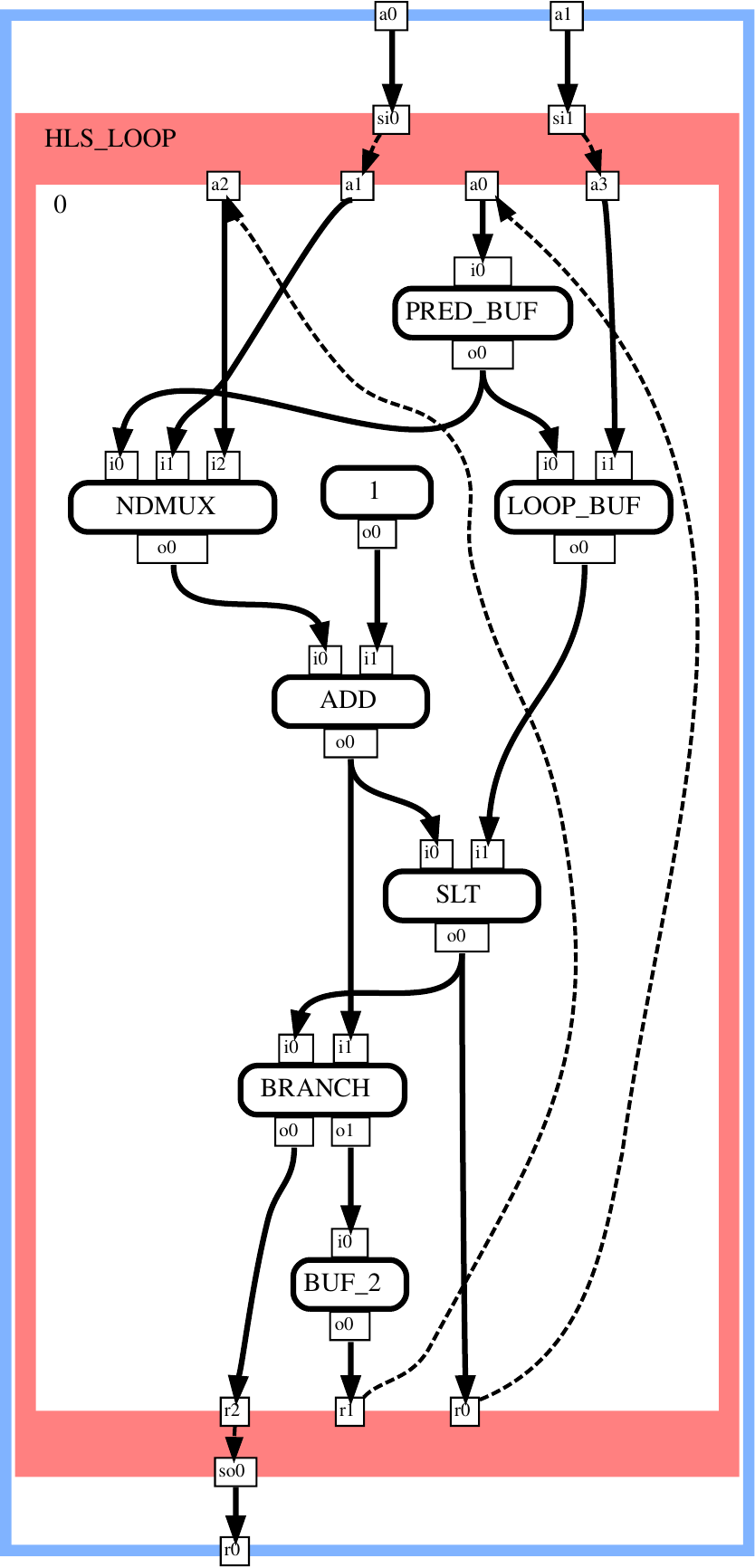}
        \caption{Lowered to \name{}}
        \label{fig:hls-loop}
    \end{subfigure}
    \vspace{-1em}
    \caption{Lowering of theta node}
\end{figure}

Theta nodes are lowered to HLS-LOOP nodes that are similar conceptually, but are more explicit with regard to their inputs and outputs.
The subregion of a theta is not flattened into the containing region to avoid cycles and to preserve clear indications of looping behavior.

For R-LLVM theta nodes, each input has an associated argument, result, and output that form a loop variable. 
An example of this are \emph{si0}, \emph{a0}, \emph{r1}, and \emph{so0} in \autoref{fig:llvm-theta}.
These maintain a one-to-one relationship, where the input gets mapped to the argument at the start of the loop.
For each new iteration, the result gets mapped to the argument via an implicit back edge, shown as a dotted line in \autoref{fig:llvm-theta}.
The back edge is implicit, since RVSDG does not allow cycles in the graph.
When the loop terminates, the result gets mapped to the output.
The only exception is the loop predicate (\emph{r0}), which controls loop continuation, and does not have a corresponding input, argument, and output.

\looseness -1
For HLS-LOOP nodes, the routing from the input or back edge is made explicit by representing each as a separate argument.
This is shown in \autoref{fig:hls-loop}, where \emph{si0} is connected to \emph{a1} and \emph{r1} to \emph{a2}.
The two arguments are in turn connected to a non-discarding multiplexer (NDMUX) that is controlled by a predicate buffer (PRED-BUF).
Similarly, the routing to either the back edge or output is made explicit by representing each as a separate result, \emph{r1} and \emph{r2}, respectively.
A BRANCH node is introduced and is connected to the two results.
%
%An example is shown in \autoref{fig:hls-loop} where \emph{si0} is connected to \emph{a1} that is connected to a non-discarding multiplexer (NDMUX), the second input of the multiplexer is connected to a new argument, \emph{a2}, that is connected through an implicit back edge (dashed line) to \emph{r1}.
%The NDMUX is controlled by a PRED\_BUF
%
%An example is shown in \autoref{fig:hls-loop}, inputs are only associated with an argument (\emph{si1} and \emph{a3}), outputs with a result (\emph{so0} and \emph{r2}) and there are result/argument pairs forming implicit back-edges (\emph{r1} and \emph{a2}).
%
%Each loop variable in a theta node is split into an input/argument, back-edge result/argument and result/output.
%For example, \emph{si0}, \emph{a0}, \emph{r1} and \emph{so0} in \autoref{fig:theta} are converted to \emph{si0}/\emph{a1}, \emph{r1}/\emph{a2} and \emph{r2}/\emph{so0} in \autoref{fig:theta_conv}.
%A non-discarding multiplexer (\emph{NDMUX}) selects between the input argument (\emph{a1}) and the back edge argument \emph{a2}.
%The multiplexer is controlled by a predicate buffer (\emph{PRED\_BUF}), 
The PRED-BUF is a special buffer type that is initialized with a loop-termination token.
At the start of the first iteration, the PRED-BUF already contains the termination token, which is driven on its output and causes the NDMUX to select the loop input, \emph{a1}.
%On each consecutive iteration, the PRED-BUF receives a loop continuation token via its back edge (\emph{a0)}, which causes the NDMUX to select its back edge (\emph{a2}).
%Once the loop terminates, the PRED-BUF receives the loop termination token, which 
%
%This means that the loop initially, and after loop termination, accepts a token from the outside and not the back-edge.
%
%A \emph{BRANCH} selects between continuing the loop by passing a token into a buffer (\emph{BUF\_2}) and then the back-edge, and ending it, by passing a token into the output result (\emph{r2}).
The PRED-BUF and BRANCH is controlled by the loop predicate, which in this example is a signed less than (SLT) operation. 
For each new iteration, the loop predicate outputs a loop-continuation token, causing the BRANCH and PRED-BUF to select their back edge \emph{r1} and \emph{a2}, respectively. 
When the loop terminates, a loop-termination token gets generated, causing the BRANCH to select the loop output,\emph{r2}, and the PRED-BUF to be initialized with a new loop-termination token, enabling a new invocation of the HLS-LOOP.
%For each new iteration, the \emph{BRANCH} routes its input token to \emph{r1} (the back edge) while it routes the token to \emph{r2} (the output) upon loop termination.

The back-edge represented by \emph{r1} and \emph{a2} would create a combinatorial loop, so a BUF is inserted after the BRANCH to break the cycle.
The PRED-BUF already breaks the combinatorial cycle for the back edge represented by \emph{r0} and \emph{a0}, so a buffer does not have to be inserted in this case.

%
% We are currently not describing how memory ports are modeled in LOOPs
%

%\hms{This should maybe be moved to a later section that explains how memory is represented, espeecially since the example does not contain this form of inputs and outputs.}
%Having inputs and outputs not directly controlled by looping behavior allows creating connections to nodes outside the loop for memory accesses.

A loop variable that does not change during the loop execution is handled separately using a loop-constant buffer (LOOP-BUF).
This is the case for the loop variable formed by \emph{si1}, \emph{a1}, \emph{r2} and \emph{so1} in \autoref{fig:llvm-theta}, which is converted into \emph{si1}, \emph{a3}, and the LOOP-BUF in \autoref{fig:hls-loop}.
When the LOOP-BUF receives a loop-termination token on its input, \emph{i0}, from the PRED-BUF, the LOOP-BUF updates its contents from \emph{i1}.
Whenever a predicate token is received, whether it is loop continuation or termination, the LOOP-BUF generates one token that replicates the contents of its buffer on its output.
While the loop constant buffer could be replaced with a non-discarding multiplexer, a branch and a buffer, the loop constant buffer consumes fewer resources, making it a superior choice.
Outputs of loop variables that remain unchanged are generally unused, since the source of the loop input can be used instead.

% \subsubsection{Local Memory}

\subsubsection{Memory Ports}
\label{sec:memory-ports}

\begin{figure}[t]
    \centering
    \begin{subfigure}[b]{0.25\textwidth}
        \centering
        \includegraphics[width=0.55\textwidth]{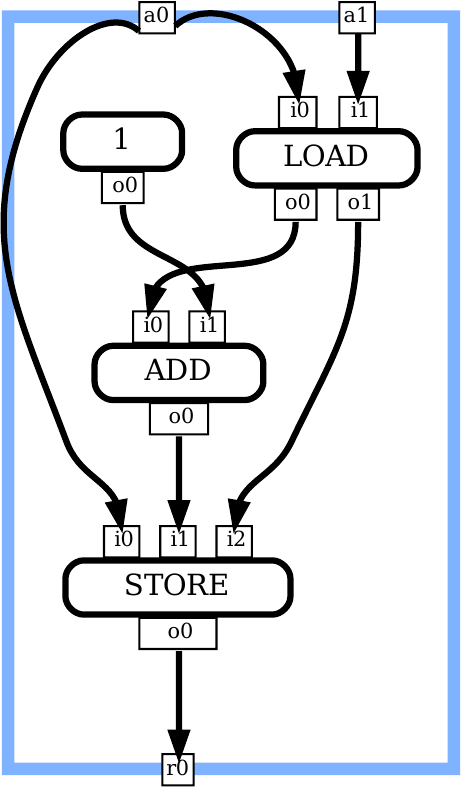}
        \caption{Memory operations in R-LLVM}
        \label{fig:mem-rvsdg}
    \end{subfigure}
    \hfill
    \begin{subfigure}[b]{0.2\textwidth}
        \centering
        \includegraphics[width=0.8\textwidth]{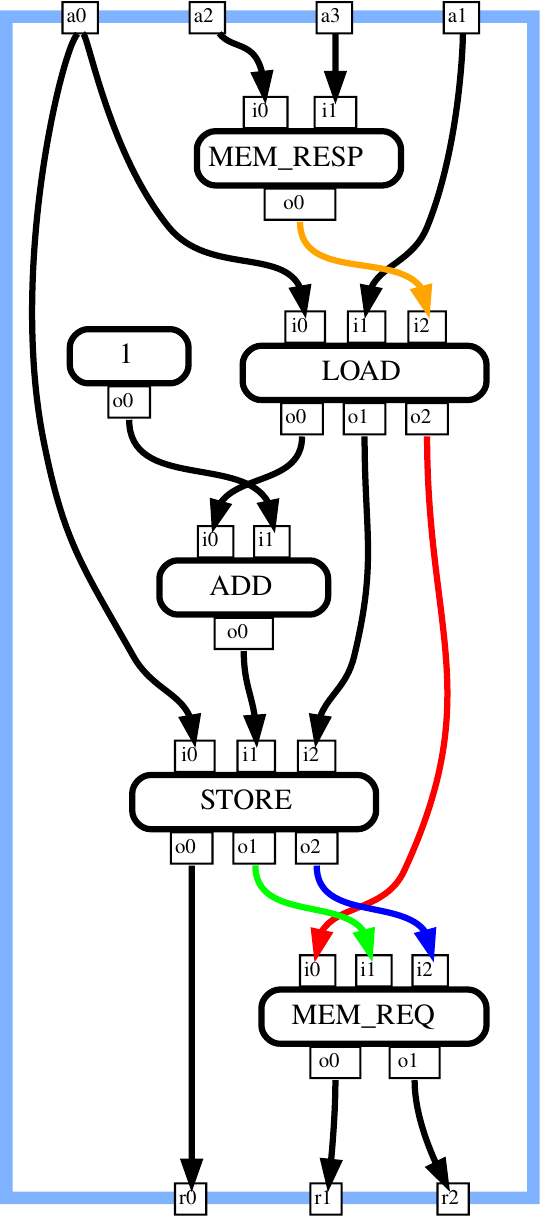}
        \caption{Lowered to \name{}}
        \label{fig:mem-rhls}
    \end{subfigure}
    \vspace{-1em}
    \caption{Memory operation conversion}
\end{figure}

Unlike software, that is provided with the illusion of one large continuous memory space, hardware accelerators need to interface with different types of memory.
This can be in the form of various local memories, such as BRAMs on FPGAs, or ports to a global memory with differing access, coherence, and caching properties.
Since R-LLVM is targeted towards optimization of software, its loads and stores are implicitly connected to global memory, as shown in \autoref{fig:mem-rvsdg}.
\name{} introduces explicit memory ports that are divided into a request (MEM-REQ) and response (MEM-RESP) portion, as shown in \autoref{fig:mem-rhls}.
Each load has an address output (red) that connects to a MEM-REQ, and a data input (orange) that receives loaded data from a MEM-RESP.
Each store has an address (green) and data (blue) output.
Currently, stores are considered to be finished the cycle after they have been sent to the MEM-REQ.
For this reason, stores are not connected to the MEM-RESP.
When combined with more advanced memory subsystems, it might be beneficial to only consider stores completed once they have received an affirmative response.
This could take the form of a state edge from the MEM-RESP to the store.

The MEM-REQ assigns a unique ID for each load and store it is connected to, and that is forwarded to the memory port.
This ID is later sent back to the MEM-RESP, which enables it to forward the response to the correct load.
For memories that support accesses of different width, the MEM-REQ provides access-width information to the port.
The MEM-REQ also handles the task of deciding which requesting memory operations are performed each cycle.

To avoid the possibility of deadlocks, which could occur if a load does not accept the response from a MEM-RESP, making other loads unable to issue requests, memory operations can only make requests that they can guarantee being able to accept a response for.
This is implemented by including a buffer inside the load and only allowing it to make requests if this buffer is empty or draining.

When the generated circuits are connected to dual-ported BRAMs, each MEM-REQ is connected to two ports with read and write capabilities.
An additional pass removes base-pointer arguments and transforms pointer calculations into index calculations.

\subsubsection{Enforcing point-to-point edges}
\label{sec:point-to-point}

Before a circuit can be generated from \name{}, all node ports need to be converted to a one-to-one relationship, i.e., each output has to have exactly one user.
This is done late in the conversion process, to avoid having to maintain the one-to-one relationships through other passes.
Each input having one defined origin node is already the case after lowering the theta and gamma.
Outputs that have no users are connected to sinks that have permanently set ready signals, i.e., always accept tokens.
A fork node is inserted for each output with multiple users.
The fork node lets each user receive a copy of a token it receives from the output, and only signals that it has consumed the token to the output once all users have done so.
Figures show R-HLS before this pass has been applied, which reduces the number of nodes and makes them easier to understand.

\subsection{Buffer Placement}
\label{sec:buffer-placement}

Circuits generated using the lowering process described above function correctly, but are only able to exploit limited parallelism due to overly tight coupling.
\name{} employs heuristics for buffer placement to alleviate this.
While more sophisticated buffer placement approaches maximizing throughput~\cite{timing-data-flow:FPL2022} 
%\david{cite https://ieeexplore.ieee.org/abstract/document/10035122} 
and optimizing circuit timing~\cite{buffer-placement:TRETS2021} 
%\david{cite https://dl.acm.org/doi/full/10.1145/3477053?sid=SCITRUS} 
have been demonstrated, the explicit nature of loops in the \name{} IR enables us to reach state-of-the-art performance, albeit at longer critical paths, without having to invest the engineering effort required to integrate them.
Therefore, advanced buffer insertion approaches are left for future work.

\looseness -1
Opaque buffers, that have no combinatorial path between their input and output, are already present to break combinatorial cycles on the back-edges of loop nodes.
To shorten the critical path, we place additional opaque buffers at the outputs of multipliers that do not have a constant input, and at the outputs of outer loops.
Multipliers have a relatively long critical path and are commonly not on the path determining the iterative intensity of a loop, enabling overlapping iterations to hide the cycle latency while maintaining throughput.
Outer loops often represent the longest paths in \name{} designs, and having two consecutive outer loops combinatorially chained is thus undesirable.
Furthermore, outer loops are only terminated once per accelerator invocation, resulting in only one cycle increase per outer loop.
Address and data outputs connecting to memory request nodes do not receive buffers, since this would worsen memory access latency and, in the case of stores, could introduce memory ordering violations.
To enable a store operation, or load operation with a state edge, to fire in consecutive cycles of a loop, the opaque buffer on the back-edge is removed if it can be directly traced to a memory operation.
In this scenario, the memory operation is responsible for splitting the combinatorial cycle along its state edge.

Transparent buffers act as FIFO queues that decouple different parts of the circuit and make it possible to exploit more parallelism through dynamic behavior.
Transparent buffers are inserted on the outputs of forks, since they represent a common point of divergence between paths of differing lengths.
Forks distributing control type tokens that control the behavior of branches and multiplexers receive larger transparent buffers, since they are relatively inexpensive due to commonly carrying only one bit of data, and determine how far ahead parts of a loop can run.

\section{Distributed Memory Disambiguation}
\label{sec:memory-disambiguation}

\begin{figure}[t]
    \centering
    \begin{subfigure}[b]{0.2\textwidth}
        \centering
        \includegraphics[width=0.9\textwidth]{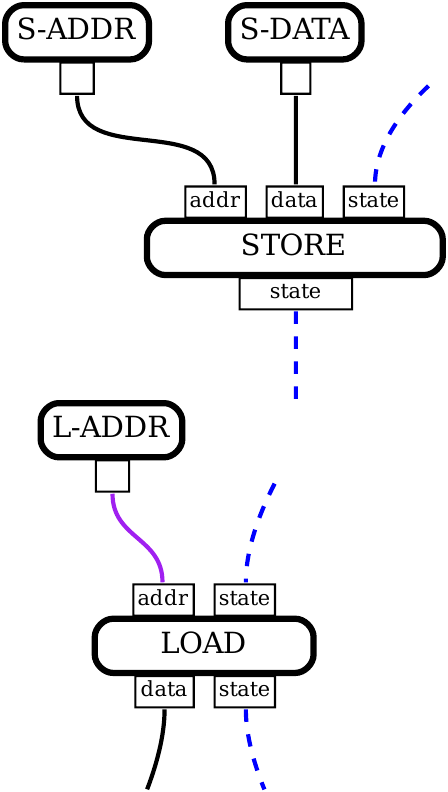}
        \caption{State-edge (dashed, blue) preserving ordering between a store and a load.}
        \label{fig:simple}
    \end{subfigure}
    \hfill
    \begin{subfigure}[b]{0.25\textwidth}
        \centering
        \includegraphics[width=0.9\textwidth]{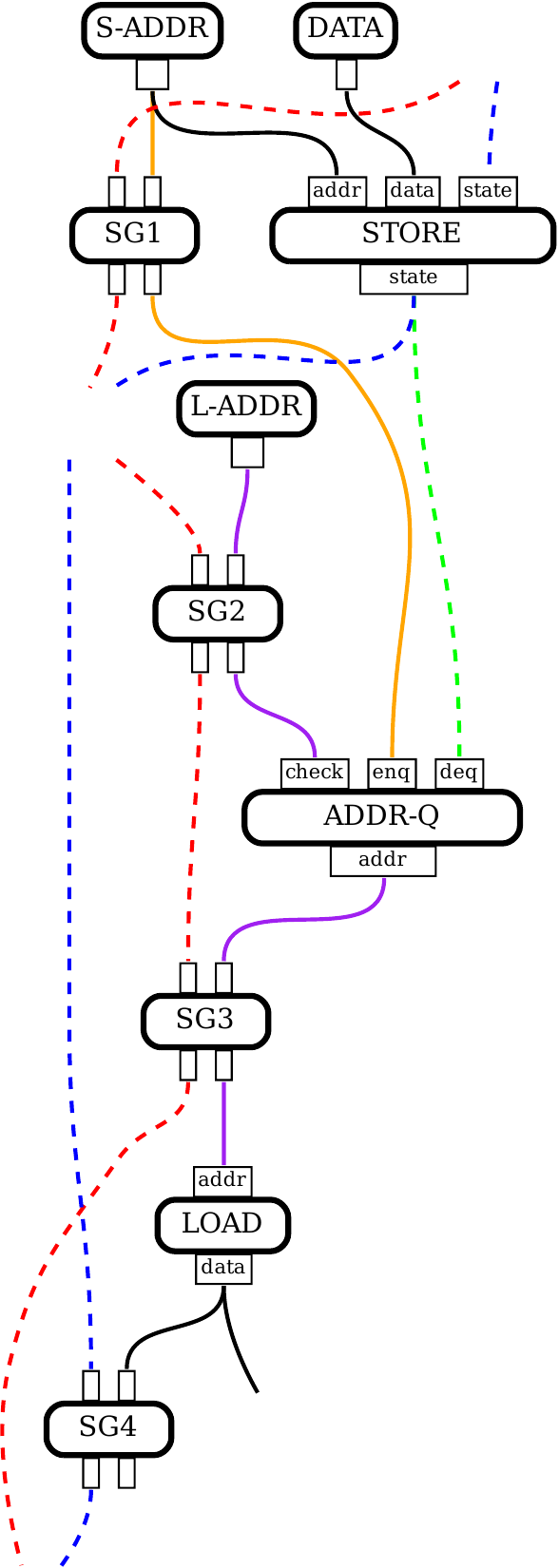}
        \caption{Distributed address disambiguation for a store and a load in a loop.}
        \label{fig:addrq}
    \end{subfigure}
    \vspace{-1em}
    \caption{Memory ordering mechanisms}
\end{figure}

By default, a load can not execute before all stores that may alias with it, and that precede it in program order, have finished executing.
In the RVSDG, this is ensured using memory state edges that turn the implicit program order into an explicit dataflow dependency, as shown in \autoref{fig:simple}.
If the load and store are part of a loop, then the state edge will be routed back up from the store to the load until the final loop iteration, where it exits the loop.
%All loads and stores that may alias are placed along one state-edge, effectively sequentializing them.
A state edge may split, to enable parallel execution of loads, and join afterward.
While the approach outlined above produces correct circuits, it is limited in performance.
In practice, while a store and a load may alias, they will often not, meaning that they could have been executed independently.
This means that, as long as address checking is performed, proving there are no uncompleted aliasing stores that precede a load in program order, the load can execute early.

Executing memory operations out of program order is a common technique for improving the performance of out-of-order (OoO) scheduled cores.
OoO cores perform memory disambiguation through the use of a load-store queue (LSQ), which keep the program order of all loads and stores and compares the address of new memory operations against older addresses stored in the LSQ.
%The LSQ is used to compare the address of a memory operation against the addresses of older memory operations.
%For example, when a new load operation is to be executed, the load address gets compared against the addresses of all older store operations.
%If all the store addresses have been computed and none of them matches the load address, then the load can be safely executed out of order, else it must wait until all older store addresses are available and none of them match.\footnote{OoO cores also speculate and execute loads ahead of stores that does not have its address computed, which results in the execution being squashed and the load to be marked to avoid further speculation if the addresses match.}
%
Memory disambiguation is crucial for the performance of dynamic OoO scheduling, and recent works have proposed the use of an LSQ~\cite{dynamically-scheduled-hls:FPGA2018, straight-to-the-queue:FPGA2023} for dynamically scheduled hardware.
%that enables memory operations to be executed out of order for dynamically scheduled hardware.
However, LSQs tend to be complex, large, and centralized hardware structures.

We propose a resource efficient and distributed memory disambiguation scheme that leverages RVSDG's memory state edges to decouple address generation and data accesses, and to dynamically compare load addresses against older store addresses.
The scheme is based on an address queue that is added for each store-load pair that may alias.
The main task of the proposed scheme is to coordinate (1) the enqueue and dequeue of store addresses to ensure that a new load address is only compared against older store addresses and (2) that writes and reads to memory are performed in correct order.
%
%younger store operations are always performed after older loads, i.e., loads can bypass older stores, but stores cannot bypass loads.
The coordination is based on the following four rules:
\\
\textit{(1) Enqueue:} A store address is enqueued as soon as an incoming address-generating state of the store operation has a token and the store address has been computed (the store operation might still be waiting for the data to be stored at this point).
\\
\textit{(2) Comparison:} Once a store address has been written to the queue, a new load address can be compared against the addresses in the queue.
If no conflict is found, then the load can proceed, potentially executing before the older store (the store might still be waiting for its data).
After the comparison, a new address token is generated, which signals that the next store address can be enqueued.
\\
\textit{(3) Dequeue:} Once a store operation completes, i.e., the data is available and written to memory, then the oldest address is dequeued.
In the case of a previous load conflict on the dequeued address, the load would have been blocked and now be able to proceed.\footnote{If the load conflicts with multiple addresses in the queue, then it would be blocked until the youngest conflicting address is dequeued, i.e., all conflicts are resolved.}
\\
\textit{(4) Store-Load Ordering}: The enqueue and comparison rule above enable the address generation of the store and load to run ahead and for younger loads to be performed ahead of older stores, as long as no address conflicts are detected.
The initial store can be performed as soon as the incoming state edge has a token, but the subsequent store can only be performed once the load has been performed and a new token is generated.
In other words, younger loads are allowed to be performed ahead of older stores, but older stores are not allowed to be performed ahead of younger loads.

\smallskip

The full implementation of the scheme is illustrated in \autoref{fig:addrq}.
%
%Address queues, that contain the addresses of uncompleted stores, are inserted in the address path of loads.
Each load has one address queue (ADDR-Q) for each store it may alias for its address path (purple).
The scheme separates the coordination of writing and reading to memory and the coordination of address generation.
To achieve this, the original state edge (dashed, blue in \autoref{fig:simple}) is duplicated into one state edge that coordinates the writing and reading of data (blue, dashed) and one state edge that coordinates the address generation (red, dashed).
%
%This queue ensures that the address of the load does not conflict with the targets of unfinished writes of the store associated to the queue.
%A second state-edge (dashed, red) is generated to follow along the original state-edge (dashed, blue).
%The blue state-edge now only controls execution of stores and not loads.
To ensure subsequent stores are not executed before a potentially conflicting load, a state-gate (\emph{SG4}) is placed along the blue state-edge and only lets a token propagate past it once the load has completed. % and outputs data.

The load is no longer controlled by a state edge and instead relies on its address input.
The ADDR-Q only lets a load address pass if it does not conflict with its contents.
The address path of the load is controlled by the red state-edge.
\emph{SG1} ensures that the state edge does not pass before the store address (orange) is enqueued, while also making sure that it is not enqueued too early in the case of loops.
\emph{SG2} ensures that the ADDR-Q is checked by a load address only after addresses of preceding stores have been enqueued into the ADDR-Q.
\emph{SG3} ensures that no new store addresses are enqueued before the address check has passed successfully.

A store address is dequeued from the ADDR-Q when the store completes.
This is accomplished using the green state edge.

This scheme might seem complex for the simple case of one load following one store, but it generalizes to multiple loads and stores in different regions.
For example, if the store is part of a loop, then multiple store addresses can be enqueued.
Once the red state edge exits the final loop operation and reaches \emph{SG2}, the load address is checked against all unfinished store addresses from the loop.

While the red and blue state edge follow the original control flow of the state edge, the enqueue (orange) and dequeue (green) signals are routed directly from the store to the ADDR-Q without having to follow control flow.
If a store precedes a load within one loop iteration, the ADDR-Q also compares the address combinatorially against the address being enqueued.
This could be avoided by placing a buffer in front of the check port, that ensures the enqueue happens at least one cycle before the check.
While a dequeue could combinatorially disable comparison against the head of the ADDR-Q, this is currently not done.

We apply this optimization separately for each outer loop, as described in \autoref{alg:addrq}.
This has the effect of sequentializing outer loops in relation to each other.
The reasoning behind not having disambiguation between outer loops is that either loop ordering does not matter (no aliasing memory operations) and they can execute independently, or ordering has to be preserved between the loops and there is only one crossing.
Since each load has an ADDR-Q for each store it may alias with, including loads and stores not contained in a loop, or in different outer loops, would be resource intensive to implement while providing limited benefits.

\begin{algorithm}[t]
\caption{Address Queue Insertion}
\label{alg:addrq}
\begin{algorithmic}[1]
\ForEach {memory state edge}
    \ForEach {outer loop along state edge}
        \ForEach {load along state edge in loop}
            \State 1: Split edge before and merge after loop
            \State 2: Follow along same control flow as original 
            \State 3: Skip other loads encountered by original edge
            \State on new edge
            \State 4: Replace load in original edge with state gate \State \emph{SG4} depending on load data output
            \State 5: Replace load in new edge with two consecutive
            \State edges \emph{SG2} and \emph{SG3}
            \State 6: Insert \emph{SG2} and \emph{SG3} into the  \State address path of load
            \ForEach {store along state edge in loop}   
                \State 1: Replace store in new edge with state gate \emph{SG1} \State dependent on store address
                \State 2: Insert address queue along load address path \State between \emph{SG2} and \emph{SG3}
                \State 3: Connect enqueue of address queue directly to \State address output of \emph{SG1}
                \State 4: Connect dequeue of address queue directly to \State state output of store
            \EndFor
        \EndFor
    \EndFor
\EndFor
\end{algorithmic}
\end{algorithm}

%\nico{The following sentence is a very strong statement. Do you have numbers to back this up?}
%\david{Yes. We have numbers for strictly enforced ordering (i.e. no dynamic disambiguation) and dynamatic uses large LSQs.}
%\nico{If that is the case, then provide a hint along the lines of: See evaluation section for details or so. I can totally see reviewers reacting to this sentence.}

Currently, \name{} generates one ADDR-Q per store-load pair in an outer loop.
For store-load pairs that seldomly conflict, or operations that rarely execute, it might be beneficial to couple the state edge controlling the load's address generation to the state output of that store, instead of having a separate ADDR-Q for it.
This, along with sizing of address queues, would provide a mechanism for more fine-grained memory dependency handling. 
Our results presented in \autoref{sec:results} show that the proposed scheme is significantly less resource-intensive to implement than a conventional load-store queue, while offering comparable performance in the common case when few memory accesses alias.

\vspace{-1em}
\section{Methodology}
\label{sec:methodology}

We implemented \name{}, as described in the previous sections, as a backend of the JLM research compiler~\cite{jlm:GITHUB2024}.
%, and intend to make \name{} publicly available once this paper is accepted.
We start with C-code that is compiled to LLVM IR using Clang release 16.
The IR is fed to the JLM optimizer, which converts the LLVM IR into an R-LLVM graph.
The R-LLVM graph is then lowered to \name{} (\autoref{sec:rllvm-to-rhls}) upon which we perform buffer placement (\autoref{sec:buffer-placement}) and our proposed memory disambiguation optimization (\autoref{sec:memory-disambiguation}).
The final \name{} graph gets lowered to the MLIR FIRRTL~\cite{firrtl:UCB2016, firrtl:ICCAD2017} Dialect, which is part of the CIRCT project~\cite{circt}.
The generated FIRRTL is finally converted to Verilog using firtool~\cite{circt}, which is simulated using Verilator and synthesized to a Kintex-7 Xilinx FPGA using Vivado.
For synthesis, we include a top-level design incorporating BRAM memories and AXI-Lite based control logic.

The generated hardware circuits are verified not only by comparing the end result against a software run execution.
We also instrument each memory operation of both the software and hardware version and validate that the corresponding reads and writes in the two versions are to the same addresses and with the same data.
This ensures that the generated hardware is functionally correct and does not create unforeseen side effects.

We use the same set of benchmarks as A. Elakhras et al. use for evaluating ``Straight to the Queue''~\cite{straight-to-the-queue:FPGA2023}, and generate results for three different circuits:
\begin{itemize}
    \item \textit{StoQ} is generated by the open-source HLS tool~\cite{straight-to-the-queue:GITHUB2024} of the ``Straight to the Queue'' work~\cite{straight-to-the-queue:FPGA2023}.  
    \item \textit{\name{}} is generated as described in the previous sections. 
    \item \textit{NoQ} is \name{} without distributed address disambiguation. 
\end{itemize}
Unlike StoQ our results include a top-level design with instantiated BRAMs.
This adds timing constraints for memory ports that are not sufficiently captured in StoQ, impacting critical paths.

%Verification: \\
%* Generate HLS + Instrumented C version, that calls on store \\
%* At runtime: \\
%    * fork process in harness to get same memory layout \\
%    * create pipe for communication \\
%    * HLS as parent, instrumented reference as child \\
%    * run both versions and compare output \\
%        * how to find first divergence? \\
%            * finish running C version first \\
%            * turn C version into address -> list of written data dictionary \\
%            * when encountering store/load check if write at head of list matches \\
%                * if not, fail \\
%        * record address ranges generated using alloca and exclude them \\

%Synthesis:
%*AXI-lite controlled top containing BRAMs
%*used for both \name and Dynamatic

\section{Results}
\label{sec:results}

\begin {table*}[t]
\caption{Results for StoQ: ``Straight to the Queue''~\cite{straight-to-the-queue:FPGA2023}, \name{}, and NoQ: \name{} without address disambiguation}
\label{tab:results}
\vspace{-2mm}
\resizebox{\linewidth}{!}{
\begin{tabular}{l|rrr|rrr|rrr|rrr|rrr|rrr}
\toprule
 & \multicolumn{3}{c}{Cycles} & \multicolumn{3}{c}{Critical Path (ns)} & \multicolumn{3}{c}{Execution Time ($\mu$s)} & \multicolumn{3}{c}{Lookup Tables} & \multicolumn{3}{c}{Flip-Flops} & \multicolumn{3}{c}{DSP Blocks} \\
 & StoQ & \name{} & NoQ &  StoQ & \name{} & NoQ & StoQ & \name{} & NoQ & StoQ & \name{} & NoQ & StoQ & \name{} & NoQ & StoQ & \name{} & NoQ \\
\midrule
2mm        &  2,498 &  2,011 &  4,009 & 7.77 &  9.16 &  8.52 &  19.41 &  18.43 &  34.15 & 22,190 &  8,174 & 6,325 &  6,715 &  7,359 & 6,230 & 12 & 18 & 18 \\
3mm        &  2,498 &  2,010 &  4,008 & 7.87 & 10.49 &  8.23 &  19.66 &  21.09 &  33.00 & 39,742 & 11,772 & 8,827 & 10,667 & 10,172 & 8,417 &  9 & 18 & 18 \\
atax       &    840 &    787 &  1,585 & 6.76 &  9.08 &  9.11 &   5.68 &   7.14 &  14.43 & 20,256 &  4,636 & 3,462 &  4,903 &  4,199 & 3,309 &  6 &  8 &  8 \\
covariance & 36,307 & 19,014 & 38,422 & 7.08 & 10.21 & 10.63 & 257.13 & 194.13 & 408.39 & 21,345 &  6,516 & 5,186 &  5,694 &  5,966 & 5,047 &  3 &  6 &  6 \\
getTanh    &  2,009 &  1,356 &  2,035 & 8.44 &  8.42 &  8.19 &  16.95 &  11.42 &  16.66 & 18,994 &  2,076 & 1,803 &  4,058 &  2,153 & 1,914 &  9 &  9 &  9 \\
histogram  &  1,016 &  1,011 &  2,005 & 6.45 &  5.78 &  5.23 &   6.55 &   5.84 &  10.49 & 19,437 &  2,306 & 1,854 &  4,198 &  2,266 & 1,858 &  0 &  0 &  0 \\
jacobi\_1d &  1,173 &    882 &  1,463 & 7.24 &  6.38 &  6.16 &   8.49 &   5.62 &   9.01 & 18,911 &  3,525 & 2,587 &  4,338 &  3,118 & 2,448 &  3 &  0 &  0 \\
triangular &  9,892 &  8,290 & 14,953 & 7.36 & 10.09 &  8.38 &  72.82 &  83.66 & 125.25 & 20,046 &  3,764 & 2,787 &  4,573 &  3,403 & 2,735 &  3 &  4 &  4 \\
\midrule
Normalized gmean & 100\% &  78\% &  145\% & 100\% &  116\% & 107\% &  100\% &  90\% &  156\% & 100\% &  21\% & 16\% &  100\% &  78\% & 65\% & 100\% & 140\%  & 140\% \\
%gmean      &  2,862 &  2,228 &  4,146 & 7.35 &  8.52 &  7.89 &  21.03 &  18.99 &  32.71 & 21,923 &  4,553 & 3,531 &  5,355 &  4,201 & 3,467 & 0 & 0  & 0 \\
\bottomrule
\end{tabular}
}
\vspace{-1em}
\end{table*}

\begin{figure}[t]
    \centering
    \includegraphics[width=\columnwidth]{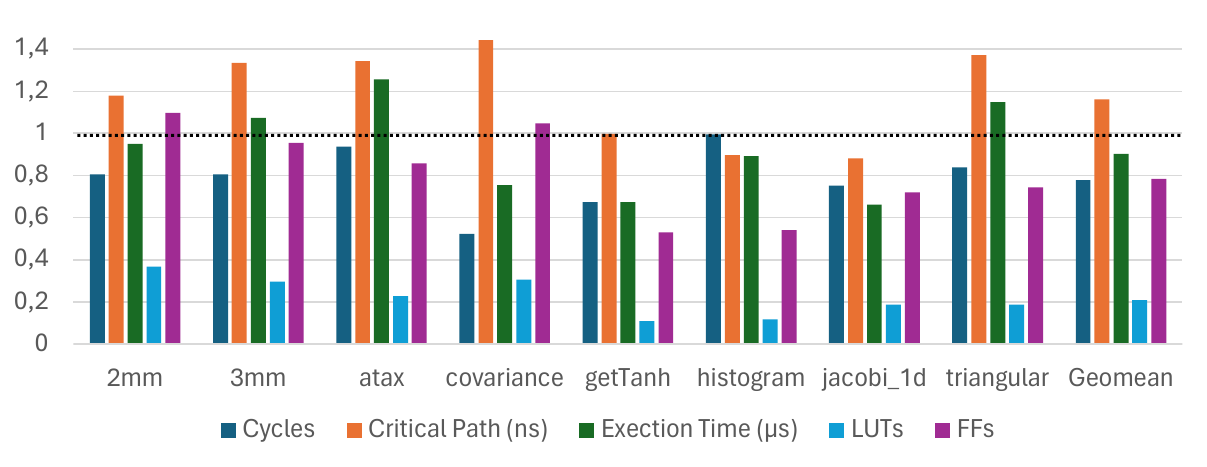}
    \vspace{-1.5em}
    \caption{Execution time and resource utilization of \name{} circuits, normalized to the results of StoQ~\cite{straight-to-the-queue:FPGA2023}}
    \label{fig:results}
    \vspace{-1em}
\end{figure}

Our results are reported in \autoref{tab:results} and \autoref{fig:results} visualizes them normalized against the previous state-of-the-art based on StoQ~\cite{straight-to-the-queue:FPGA2023}.
%We evaluate and compare our circuits using the benchmarks used in~\cite{straight-to-the-queue:FPGA2023} and include a variant of \name{} without distributed address disambiguation (\name{}-NQ), that sequentializes memory operations along state edges.
We have a geometric mean (geomean) execution time reduction of 10\% compared to StoQ~\cite{straight-to-the-queue:FPGA2023} with at best a 33\% execution time reduction and at worst a 26\% execution time increase.
The execution time is a result of a geomean decrease in cycle count of 22\%, offsetting a 16\% increase in critical path length.
LUT utilization is consistently much lower for \name{}, with it using between 89\% and 63\% less, with a geomean of 79\%.
FF utilization ranges between 10\% more and 47\% less, with a geomean of 22\%.
DSP utilization is higher for \name{}, except for jacobi\_1d.

For the critical path, and thus to some extent execution time, StoQ~\cite{straight-to-the-queue:FPGA2023} benefits from Dynamatic's more mature infrastructure that includes timing based buffer placement and sizing using MLIP solvers~\cite{timing-data-flow:FPL2022, buffer-placement:TRETS2021} compared to our relatively simplistic heuristic based approach (\autoref{sec:buffer-placement}).

Since we currently do not collect buffer occupation statistics to guide buffer placement, many buffers are likely to be over-sized.
We prioritized low cycle times, since the cycle time being consistently lower demonstrates \name{}'s suitability for extracting parallelism, as well as the effectiveness of our memory disambiguation.
Flip-flop utilization is also largely determined by buffers, and while \name{} is ahead on average, utilization could be reduced with an improved buffer placement strategy.
\name{} would also benefit from a strength reduction pass aimed at hardware generation to reduce DSP usage, and a width inference pass to reduce bit-width of operations and signals and thus reduce the general hardware utilization.

\looseness -1
Covariance is the benchmark with the highest increase in the critical path for \name{}.
The critical path starts and ends at the data output and input of a BRAM respectively, meaning that it is impacted by the inclusion of our top-level design, which includes BRAMs.
StoQ's synthesis methodology would not capture this path, since it does not contain a top-level design, and therefore does not include timing paths with BRAMs~\cite{straight-to-the-queue:FPGA2023}.
The path is between two loop bodies and could easily be divided up with better buffer placement, which by our estimate would add 32 cycles to the cycle count of 19,014.

\looseness -1
Compared to NoQ, \name{} uses a geomean of 31\% more LUTs and 20\% more flip-flops, resulting in an 9\% longer critical path.
This is however offset by a 46\% geomean reduction in the number of cycles, resulting in a 42\% execution time reduction.
This quantifies the cost of our distributed memory disambiguation scheme, and demonstrates its effectiveness at de-sequentializing memory operations.

\name{} consistently produce circuits with a reduction in cycle counts and with significantly reduced resource utilization.
Current limitations in buffer placement (\autoref{sec:buffer-placement}) result in the critical path for five of the eight benchmarks being longer.
\name{} is still 10\% faster on average, and improved buffer placement heuristics are likely to improve the critical path without significantly impacting the cycle count.

\vspace{-0.5em}
\section{Related Work}
\label{sec:related-work}

To our knowledge, Dynamatic~\cite{dynamatic:FPGA2020} is the only work that generates data flow circuits that support dynamic memory dependency resolution.
Several works build upon Dynamatic and its LSQs, focusing for example on LSQ sizing~\cite{lsq-sizing:ICFPT2022} or reducing LSQ usage~\cite{Josipovic:ICFPT2019}.
Dynamatic utilizes a basic block focused and controlled data flow, but there are works that enable the execution of multiple basic blocks at the same time~\cite{inter-block-scheduling:FPL2022, parallel-control-flow:TRETS2023}, or moving away from basic blocks as the unit of control~\cite{fast-tokens:FPL2022, straight-to-the-queue:FPGA2023}.
All of these works have in common that their analyses and transformations focus on basic blocks in a control flow graph, and that they use LSQs for memory dependency handling that can not be decided statically.
The concept of state edges is used for scheduling by the value state flow graph~\cite{vsfg:TRETS2015}, but this work sequentializes memory operations, limiting performance.

In contrast, \name{} utilizes a global data flow graph that encodes memory dependencies as state edges.
The state edges enable address generation management and disambiguation at the per store-load pair granularity, instead of at the basic block level.
This enables fine-grained analyses, optimization, and creation of distributed and resource efficient  memory disambiguation.

%There have been several intermediate representations for optimization and generating proposed, many of them based on the MLIR~\cite{mlir:CGO2021} compiler framework, among them $\mu$IR~\cite{muIR:MICRO2019} and ScaleHLS~\cite{scalehls:HPCA2022}.

% Pegasus~\cite{pegasus:REPORT2002} \\
% ``We propose an alternative --- decouple the representation used for accelerator microarchitecture and hardware optimizations from the functional behavior specification.'' \\
% ``$\mu$IR is a structural graph that explicitly specifies the accelerator’s microarchitecture components and orchestrates data movement between the different components.'' \\
% C-to-RTL~\cite{c-to-rtl:DSD2010} \\
% From functional programs to dataflow circuits~\cite{from-functional-to-dataflow:CC2017} \\
% Static Tokens: Using Dataflow to Automate Concurrent Pipeline Synthesis~\cite{static-tokens:ASYNC2004} \\
% Intermediate representations survey~\cite{ir:CSUR2013} \\
% OoO-LQ~\cite{ooo-lq:TECS2017}: \\

\vspace{-0.5em}
\section{Conclusion}

\name{} demonstrates the suitability of region-based data flow intermediate representations for dynamic high-level synthesis and champions an alternative to expensive centralized load-store queues in the form of distributed memory disambiguation based on state edges.
\name{} integrates distributed dynamic memory dependency handling as both part of the circuit and the intermediate representation, and enables fine-grained trade-offs per memory operation pair.
\name{} enables a geomean execution time reduction of 10\% compared to the state-of-the-art, while consuming 79\% fewer LUTs and 22\% fewer flip-flops.

\vspace{-0.5em}
\section*{Acknowledgement}

\name{} has partly been developed on the IDUN/EPIC~\cite{epic:ARXIV2022} computing cluster at NTNU.

\balance 
\bibliographystyle{ACM-Reference-Format}
\bibliography{refs}

\end{document}